\documentclass[preprint]{aastex62}

\usepackage{mathtools}
\usepackage{lineno}
\usepackage{mathrsfs}

\submitjournal{ApJ}

\shorttitle{[FUV-MUV] color and Mg II index variations}
\shortauthors{Criscuoli et al.}

\begin{document}
\title{The correlation of synthetic UV color vs Mg II index along the solar cycle}

\correspondingauthor{Serena Criscuoli}
\email{scriscuo@nso.edu}

\author[0000-0002-4525-9038]{Serena Criscuoli}
\affil{National Solar Observatory\\
3665 Discovery Dr. \\
Boulder, CO 80303, USA}

\author{Valentina Penza}
\affiliation{University of Roma “Tor Vergata” \\
Via della Ricerca Scientifica 1,  \\
I-00133 Roma, Italy}

\author{Mija Lovric}
\affiliation{University of Roma “Tor Vergata” \\
Via della Ricerca Scientifica 1,  \\
I-00133 Roma, Italy}

\author{Francesco Berrilli}
\affiliation{University of Roma “Tor Vergata” \\
Via della Ricerca Scientifica 1,  \
I-00133 Roma, Italy}
 
\begin{abstract}
UV solar irradiance  strongly affects the chemical and physical properties of the Earth's atmosphere.
UV radiation is also a fundamental input for modeling the habitable zones of stars and the atmospheres of their exo-planets. 
Unfortunately,  measurements of solar irradiance are affected by instrumental degradation and are not available before 1978. For other stars, the situation is worsened by interstellar medium absorption. 
Therefore, estimates of solar and stellar UV radiation and variability often rely on modeling.
Recently, \citet{lovric2017} used SORCE/SOLSTICE data to investigate the variability of a color-index that is a descriptor of the UV radiation that modulates the photochemistry of planets atmospheres.
After correcting the SOLSTICE data for residual instrumental effects, the authors found the color-index to be strongly correlated with the Mg II index,
 a solar activity proxy.
In this paper we employ an irradiance reconstruction to synthetize the UV color and Mg II index with the purpose of investigating the physical mechanisms that produce the strong correlation between the  color-index and the solar activity.
Our reconstruction, which extends back to 1989, reproduces very well the observations, and shows that the two indices can be described by the same 
linear relation for almost three cycles, thus ruling out an overcompensation of SORCE/SOLTICE data in the analysis of \citet{lovric2017}. We suggest that the strong correlation between the indices results from the UV radiation analyzed 
originating in the chromosphere, where atmosphere models of quiet and magnetic features present similar temperature and density gradients. 
\end{abstract}

\section{Introduction}
Variations of Total Solar Irradiance (TSI) and Spectral Solar Irradiance (SSI) measured at different temporal scales affect the Earth's chemistry and dynamics, thus affecting
the Earth's climate  \citep[e.g.][]{matthes2006,gray2010,lockwood2011,ermolli2013,seppala2014,schwander2017, matthes2017}. 
It is well known that most of these variations are modulated by the magnetic activity, in particular by the appeareance on the solar surface of magnetic structures as plages and sunspots, 
which, in turn, modify the radiative properties of the solar atmosphere. 
The principal temporal scale of variability is the 11-year cycle, observed since the 17th century, along which both, the number
of active regions and the total and spectral intensity vary.

Of particular importance are variations measured at wavelengths shorter than approximately 400 nm, which is mostly absorbed by the high and medium layers of the Earth atmosphere. In particular, solar radiation 
at wavelengths shorter than 120 nm (Extreme UV) plays an important role in creating the Earth ionosphere.  Irradiance variations at these wavelengths are responsible for variations of geomagnetic activity, may 
create disturbances in radio wave propagations and may modify satellite orbits due to the air-drug increase \citep[e.g.][]{floyd2002}. In the stratosphere, heating is caused by direct absorption of near-UV radiation
in the  Hartley (200-300 nm) and Huggins (320-360 nm) bands,  whereas ozone photo-dissociation peaks in the Hartley continuum, at about 250 nm \citep{haigh1994}.  Variations of ozone abundance, in turn, cause further 
heating which, through the top-down mechanism, changes the atmospheric dynamics, thus affecting the tropospheric circulation patterns \citep[][]{scaife2013,bordi2015,gray2016,matthes2017}.

Variations of solar irradiance in the UV have been continuously monitored with various instruments since the launch of NIMBUS-7 in 1978 \citep[see for instance][for a review]{deland2012}, whereas continuous
monitoring of the EUV started only in 2002 with the launch of the Thermosphere Ionosphere Mesosphere Energetic and Dynamics (TIMED). Unfortunately, measurements of solar irradiance are known to be affected by
long-term instrumental degradation effects, so that while variations obtained with different instruments typically agree when compared on temporal scales of the order of a few solar rotations, large differences are 
found at the solar cycle and longer temporal scales \citep[e.g.][]{deland2012,ermolli2013,yeo2014}. Such discrepancies hinder estimates of the impact of solar activity on the Earth atmosphere properties and 
in particular of the ozone production \citep[e.g.][]{merkel2011,dhomse2016,matthes2017}.

In order to provide the community with consistent spectra over long temporal scales (decades), and to extend them to times where measurements are not available, TSI and SSI measurements are complemented with 
independent estimates obtained with semi-empirical approaches \cite[e.g.][]{penza2006,fontenla2011,ermolli2011,yeo2014} and/or through regression analyses with proper activity indices 
\citep[e.g.][]{dudokdewit2009,bolduc2012,thuillier2012,coddington2016}.

In the recent past, studies of solar variability have been also motivated by the necessity of improving our understanding of stellar variability \citep[e.g.][for a recent review]{fabbian2017}, with the aim of characterizing
the habitable zones of stars and the atmospheres of their exo-planets. Likewise the Earth's atmosphere, modeling of exoplanets requires as fundamental input the radiation emitted by the host star in the UV and shorter 
bands \citep[e.g.][]{tian2014,madhusudhan2014,shields2016,omalley2017}.  For this reason, large efforts have been dedicated to measure stellar UV and XUV spectra \citep[see][for a recent review]{linsky2017}, with recent 
particular interest in spectra of K and M dwarfs \citep[e.g.][]{france2016,youngblood2016}, their planets being suitable for spectroscopic biomarker searches \citep[e.g.][]{scalo2007,cowan2015,kaltenegger2017}.

In the case of stellar fluxes, measurements in the FUV and shorter wavelengths are strongly hampered by interstellar medium absorption, which is significant even for relatively close stars. 
Consequently, estimates of spectra at these spectral ranges have to rely on models \citep[e.g.][]{wood2005,youngblood2016, fontenla2016} or proxies \citep[e.g.][]{linsky2013,shkolnik2014,smith2017}. It is important
to note that there is no space mission scheduled in the near future aimed at observing stellar spectra in the UV and EUV ranges, so that after the Hubble Space Telescope will cease operations, estimates
of spectra at short wavelengths will necessarily rely on the use of models and/or proxies measured at longer spectral ranges.

Within this framework, \citet{lovric2017} introduced a spectral color index  in the solar UV that is linked to the ratio between the flux integrate over the Far-UV and Middle-UV spectral broad-bands. 
Such a descriptor can be used to characterize UV stellar emission which modulates the photo-chemistry of molecular species, e.g., oxygen, in the atmospheres of planets \citep[e.g.][]{tian2014}. 
By using solar irradiance measurements obtained with radiometers aboard the Solar Radiation and Climate Experiment satellite \citep[SORCE,][]{ mcclintock2005} for almost a solar cycle, the authors showed that the 
color index so defined is linearly correlated with the Bremen Magnesium II index, which is an excellent proxy of the magnetic activity \citep{viereck2004}.  Lovirc et a. showed that the correlation coefficient is slightly different 
for the descending phase of Cycle 23 and the 
ascending phase of the subsequent cycle. Such difference was ascribed to residual instrumental effects, which, when compensated for, lead to a  correlation coefficient constant with time. On the other hand, several solar photospheric and chromospheric indices present a clear asymmetry during different phases of a cycle \citep[namely an hysteresis pattern, e.g.][]{bachmann1994,criscuoli2016,salabert2017}, whereas correlation coefficients between indices may vary from cycle to cycle \citep[e.g.][]{bruevich2014,tapping2017}, so that the 
question arises whether, and to what extent, the data corrections applied by \citet{lovric2017} might include a physical variation of solar emission. 
The purposes of this paper are to answer this question and to investigate the physical mechanisms determining the high linear relation between the UV spectral color index and solar activity measured with the Mg II index. 
To this aim  we compare the results of \citet{lovric2017} with synthetic indices obtained with an irradiance reconstruction technique based on the use of semi-empirical atmosphere models and full-disk observations.
%This approach is similar to other ones previously presented in the literature, as SATIRE \citep{unruh2000,krivova2006}, COSI \citep{shapiro2010} and SRPM \citep{fontenla2011,fontenla2015},
%OAR \cite{penza2004a,penza2004b,penza2006}. 

The paper is organized as follows. In Sec.~\ref{sec:data} we describe the solar irradiance measurements analyzed, we describe the UV color index and briefly summarize the correction procedure applied to the data. 
In Sec.~\ref{sec:reconstruction} we describe in detail the
irradiance reconstruction technique, and the input data and the radiative transfer utilized. Results are presented in Sec.~\ref{sec:result} and discussed in Sec.\~ref{sec:disc}. Finally, our conclusions are drawn in Sec.~\ref{sec:concl}.

\section{The solar UV color index and solar data analysis}\label{sec:data}
Our investigation is based on the analysis of two solar datasets. The first consists of  Far-UV (FUV, 115-180 nm) and Middle-UV (MUV, 180-310 nm) spectral irradiance observations acquired with the Solar 
Stellar Irradiance Comparison Experiment (SOLSTICE) \footnote{available at http://lasp.colorado.edu/lisird/} aboard the SORCE satellite, for the period May 2003 - January 2015. The second  is the Bremen 
composite of the Magnesium II core-to-wing ratio, i.e., Mg II index \citep{viereck2004} \footnote{available at http://www.iup.uni-bremen.de/UVSAT/Datasets/mgii}.
We provide here a basic description of the datasets, their treatment and a definition of the color index. An extensive description of the datasets, data analysis and correlations with solar magnetic activity
of the computed indices is provided in \citet{lovric2017}. \\
Both SOLSTICE  FUV and MUV spectral irradiance observations  were  interpolated using a Piecewise Cubic Hermite Interpolating Polynomial scheme to take into account of missing data.
Fluxes  in each band were then integrated using a Riemann integration algorithm, and finally the monthly averages of FUV and MUV spectral fluxes were computed.
The fluxes were used to derive the FUV and MUV magnitudes necessary to compute the [FUV-MUV] color index,  defined as:
\[[FUV- MUV] =  - 2.5 \cdot log \frac{F_{FUV}}{F_{MUV}}\]
Where $F_{FUV}$ and $F_{MUV}$ are the fluxes integrated in the corresponding bands. The zero points usually present in the color definition are here arbitrarily set to zero corresponding to a simple constant
offset in the final magnitude difference. Results presented in \citet{lovric2017} show that the [FUV-MUV] color index is highly correlated with the Mg II index,  indicating that the ratio of the fluxes in FUV 
and MUV bands is proportional to the solar activity on the time scale of 11 years. Nevertheless, the  correlation between the [FUV-MUV] color and the Mg II index shows slightly different coefficients for the 
descending phase of Cycle 23 and the ascending phase of Cycle 24 \citep[as shown in][ Fig.~1]{lovric2017}. \\
The observed  [FUV-MUV] color vs Mg II index correlation  can be completely reproduced by a simple model including appropriate color temperature  values changing during the solar activity cycle and a suitable
UV instrumental degradation. Consequently, this model was only used to estimate the possible instrument degradation effect in the FUV SOLSTICE fluxes able to reproduce the measured slope difference during 
the descending and the ascending phase of solar cycle. Assuming the same degradation process for all wavelengths in the FUV  an average reduction of about 0.0002\% efficiency per month is enough to remove 
the observed slope difference \citep{lovric2017}. Under this hypothesis the [FUV-MUV] color index shows the same linear relation over the whole analyzed  temporal range.

\section{Synthetic reconstruction of UV color and Mg II index}\label{sec:reconstruction}
Irradiance variations of the FUV and MUV bands were reconstructed using a semi-empirical approach similar to the one described in \citet{penza2003,penza2006},
which is based on the widely accepted assumption that irradiance variations observed at temporal scales from days to decades are modulated by variations of surface magnetism. 
In summary, spectra synthetized in Local Thermodynamic Equilibrium (LTE) are generated using semi-empirical atmosphere models representing archetypal magnetic and quiet structures. 
These are then weighed with area coverages of magnetic features as derived from full-disk CaIIK and red photospheric continua.  
The variations of the Mg II index are estimated in the same way, with the main difference of using a Non Local Thermodynamic Equilibrium (NLTE) synthesis. 
The spectral synthesis, data employed and procedure are describe in more detail below. 

\subsection{Atmosphere models}
\label{sec:atmos}
Various atmosphere models have been presented in the literature to describe solar and stellar atmospheres \citep[e.g.][for a recent review about chromospheric models]{linsky2017}. 
For our reconstructions we employed the set of seven semi-empirical
atmosphere models from the Solar Irradiance Physical Modeling (SRPM) system described in \citet{fontenla2011} (FAL2011 hereafter). The different models correspond to different types of quiet and
 magnetic features, and consist of two quiet models (model 1000 and model 1001), two network models (model 1002 and model 1003), two facular models (model 1004 and model 1005), 
 and an umbral (model 1006) and a penumbral (model 1007) model. 

%%%%%%%%%%%%%%%%%%%%%%%%%%%%%
%\begin{figure}
%    \includegraphics[width=7cm,height=6cm]{Temperatures.eps}
   
% \caption{Temperature variation with height for model 1001. Black: original model truncated at 1500 km. Red: model modified not to have a chromospheric temperature rise.}
%  \label{temps}
%\end{figure}
%%%%%%%%%%%%%%%%%%%%%%%%%%%%%

To avoid unrealistic chromospheric and coronal line emissions that would result from the LTE synthesis in the FUV and  MUV spectral ranges,  
the atmosphere models were truncated at about 1500 km above the photosphere, which corresponds roughly to the base of the transition region \citep[c.f.r. Fig.~1 in][]{fontenla2011}. 
This approximation is also motivated by previous studies showing that
the majority of the radiation in these bands forms in the higher photosphere and chromosphere \citep{thuillier2012}; in particular, only continua at wavelengths shorter than approximately 160 nm have appreciable contribution 
above the temperature minimum 
\citep[e.g.][]{uitenbroek2004, fossum2004, fontenla2015}.
The modification of atmosphere models for LTE synthesis in the UV is rather common, especially in irradiance reconstruction techniques based on LTE synthesis, although typically the modification 
is performed by linearly extrapolating the atmosphere beyond the
temperature minimum, thus replacing the chromospheric temperature rise with a monotonic temperature decrease. This is for instance the approach adopted in the Spectral And Total Irradiance REconstructions 
\citep[SATIRE,][]{unruh1999,krivova2006} and in \citet{penza2003, penza2004a}.
As very well explained in \citet{krivova2006}, while this approximation allows to reproduce relatively well the observed solar spectrum in the MUV, it produces  discrepancies at shorter wavelengths,
especially in the FUV, where the spectrum is largely underestimated. To improve the synthesis in this spectral range, in SATIRE
the synthetic spectrum is multiplied  by correction coefficients derived by the observed spectrum; old versions of SATIRE employed SUSIM data \citep{krivova2006}, while more recent versions employ the "Solar Spectral Irradiance Reference Spectra for Whole Heliosphere
Interval (WHI) 2008" \citep{woods2009} and SORCE/SOLSTICE observations \citep{yeo2014b}.
Because one of the objectives of our analysis is to verify the goodness of the instrumental degradation correction proposed in \citet{lovric2017}, and verify that this does not overcompensate for 'real' solar trends,
we decided not to apply any correction factor derived from observations to our synthetic FUV spectrum. The choice of truncating the atmosphere at 1500 km in height is in this respect a 'compromise' that allows to avoid unrealistic line emission, especially in the FUV, 
while reducing the discrepancies between the observed and synthetic spectrum.

\subsection{Synthesis of MUV and FUV spectrum}
\begin{figure}
\centering
     \includegraphics[width=8cm]{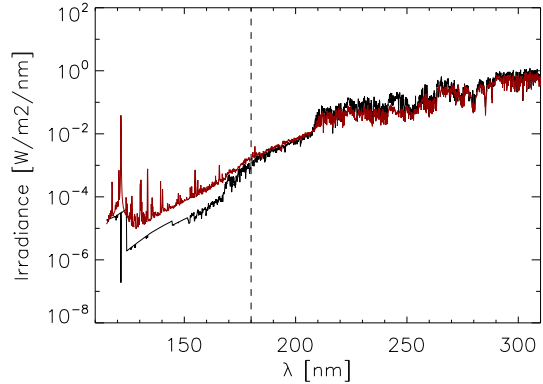}
  \caption{Comparison of the spectral synthesis (black) in the FUV (left of the vertical dashed line)  and MUV (right of the vertical dashed line) spectral ranges with the WHI Reference Spectrum (red). The irradiance scale is logarithmic. 
  Synthetic spectra were degraded to the spectral resolution of the observations.}
  \label{MUVFUV_vs_WHI}
\end{figure}

Numerical computation of stellar atmosphere opacities strongly relies on approximations and is affected by our limited knowledge of atomic and molecular parameters \citep{allendeprieto2009}. The discrepancies with observations
are enhanced in NLTE computations, which produce over-ionization of metals and a 
consequent underestimate of line and continuum opacities. The problem is particularly evident at wavelengths shorter than about 300 nm \citep[e.g.][]{bruls1992, busa2001, allendeprieto2003,shapiro2010}, most likely 
due to uncertainties in  estimates of atomic and molecular photolysis (photo-ionization and photo-dissociation) parameters. As a consequence, LTE computations may perform better in reproducing the UV spectrum than NLTE 
ones \citep[e.g.][]{allendeprieto2003,shapiro2010}. 
For our computations of the MUV and FUV spectral ranges we therefore decided to employ an LTE approach. To this aim, we used the SPECTRUM code, written by R.O. Gray. It is a stellar spectral synthesis program, written 
in the C language and distributed with an atomic and molecular line list for the optical spectral region from 90 nm to 4000 nm. It uses as input the run of the temperature structure and the electronic density, computing by 
itself the pressure and the density of the more important species. 
Here we use the latest version, SPECTRUM 2.76 \footnote{The code and the manual are available at  http://www.appstate.edu/~grayro/spectrum/spectrum.html.}, but good results were also obtained with older versions 
\citep[e.g.][]{penza2004a, penza2004b,penza2006}. As explained in Sec. \ref{sec:atmos}, to avoid unrealistic emission of chromospheric and coronal lines, especially in the FUV, which would result from the LTE treatment, 
the models were truncated at approximately 1500 km, 
corresponding to the end of the temperature plateau. 
The LTE treatment in the UV with models with chromospheric rise is a critical point of our analysis and is further discussed in Sec.~\ref{sec:disc}. Here we note that, overall, our synthesis seems to reproduce reasonably
well the observed spectrum, as shown in Fig.~\ref{MUVFUV_vs_WHI}. In particular, the plot shows the synthetized spectrum using quiet sun model 1001 and the WHI observed spectrum. The agreement is obviously more satisfactory 
in the MUV than in the FUV, where emission lines, especially the strong hydrogen Ly$\alpha$ line, are not reproduced. On the other hand, we note that that the slope of the continuum intensity is roughly reproduced,
although the flux is underestimated at wavelengths shorter than approximately 170 nm.

%%%%%%%%%%%%%
%\begin{figure}
%    \includegraphics[width=10cm,height=10 cm]{MUV_vs_WHI2.eps}
%  \caption{Comparison of the spectral synthesis in MUV range with WHI Reference }
%  \label{MUV_vs_WHI}
%\end{figure}
%%%%%%%%%%%%%
%\begin{figure}
%    \includegraphics[width=10cm,height=10cm]{FUV_vs_WHI.eps}
%  \caption{Comparison of the spectral synthesis in FUV range with WHI Reference. The irradiance scale is logarithmic. }
%  \label{FUV_vs_WHI}
%\end{figure}
%%%%%%%%%%%%%%%%%%%%%%%%%%%%%%

%%%%%%%%%%%%%
%%%%%%%%%%%%%%%%%%%%%%%%%%%%%
\subsection{Synthesis of the Mg II line}
\begin{figure}
\centering
\includegraphics[width=8cm]{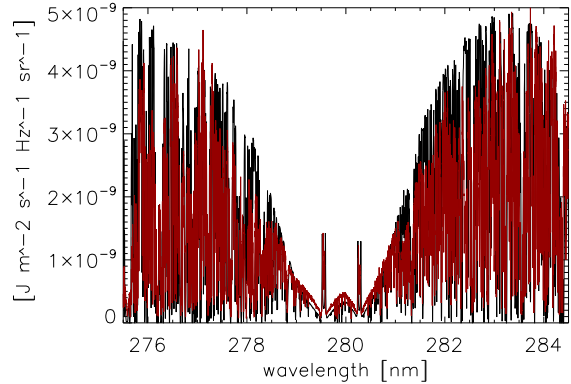}
\caption{\label{FigMgRH_vs_meas}Comparison of Mg II line spectral synthesis obtained with model 1001 at disk center (black) and the Hawaii UV Atlas (red). Synthetic spectrum was convolved with a 0.0014 nm width Gaussian function to match 
the resolution of the observations.}
\end{figure}

The Mg II h\&k lines were synthetized in NLTE \citep[e.g.][]{milkey1974,ayres1976, linsky1985,sukhorukov2017} with the RH code \citep{uitenbroek2001} together with the original (i.e. not-truncated) FAL2011 models described above.
The adopted model atom, energy levels, lines, photo-ionization and collisional parameters are described in \citet{leenaarts2013a}. Lines in the spectral range 275-286 nm from the Kurucz database where included
in the computations using a LTE approximation. As describe in previous section, the continuum opacity is underestimated at these wavelength ranges. To improve the agreement with the observed spectrum,
the continuum opacity was therefore multiplied by 
correction factors as described in \citet{bruls1992}. The comparison of the 
synthetic spectrum obtained with model 1001 at disk center with the Hawaii UV Atlas \citep{allen1977} measurements is shown in Fig.~\ref{FigMgRH_vs_meas}. The agreement is excellent in the core, while in the wings the agreement is less
satisfactory. The residual difference has to be ascribed to uncertainties in the temperature stratification of the model as well as 'missing' line opacity \citep{busa2001, collet2005, fontenla2009}.    

%Figure \ref{FigMgRH_vs_meas} shows the
%comparison of our synthetic spectrum  obtained at disk center with model 1001 and t. We note an excellent agreement in the core, whereas the agreement in the wings is not as good. 

%while bottom panel shows a comparison of the synthetic irradiance with the "Solar Spectral Irradiance Reference Spectra for 
%Whole Heliosphere Interval (WHI) 2008" \citep{woods2009}. Both plots show an excellent agreement in the line %core, and an excess of flux in the wings, especially in the red one. 
%The opacity fudge for continuum is justified by the empirical evidence that relative differences between spectra observed in  solar and facular regions are much larger in the core rather than in the wings (\cite{morrill2001})

%%%%%%%%%%%%%%%%%%%%%%%%%%%%%
%\begin{figure}
%    \includegraphics[width=10cm,height=10cm]{MUV_FAL.eps}
%  \caption{ Correlation between the normalized MUV values and Mg II index: the crosses are the values corresponding to different FAL models (from mod 1000 to mod 1005), while  the red points are the measured variations along the solar cycle.}
%  \label{MUV_FAL}
%\end{figure}
%%%%%%%%%%%%%%%%%%%%%%%%%%%%%

%%%%%%%%%%%%%%%%%%%%%%%%%%%%%
%\begin{figure}
%    \includegraphics[width=10cm,height=10cm]{FUV_FAL.eps}
%  \caption{ Correlation between the normalized FUV values and Mg II index: the crosses are the values corresponding to different FAL models (from mod 1000 to mod 1005), while  the red points are the measured variations along the solar cycle.}
%  \label{FUV_FAL}
%\end{figure}
%%%%%%%%%%%%%%%%%%%%%%%%%%%%%

\subsection{Synthetic reconstruction of UV spectrum variability}
Radiance variations were estimated by weighing the synthetic spectra obtained for the different atmosphere models 
 with corresponding coverage factors \citep[c.f.r.][]{penza2003,penza2006,ermolli2011,fontenla2011,yeo2014} as:
%%%%%%%%%%%%%%%%%%%%%
\begin{equation}
 \label{F_tot}
F(\lambda) =  \sum_{j} \alpha_{j} F_{j}(\lambda),
\end{equation}
%%%%%%%%%%%%%%%%%%%%%%%
 where $\alpha_j$ is the coverage factor corresponding to the j-th structure (quiet sun, network, facula or spot umbra and spot penumbra) and $F_{j}(\lambda)$ is the corresponding flux.
 In this model we suppose that the coverage factors are independent of disk position and use the disk integrated  fluxes. In \cite{penza2006} is shown that the error in neglecting the disk-position of magnetic features is negligible.   
We are able to reconstruct the UV variations over the time range from 1988 to 2015 by utilizing magnetic features area coverages obtained from two different sets of full-disk observations. 
From 1988 to 2004 we use area coverages of sunspot and faculae obtained at the San Fernando Observatory 
\citep{walton2003,preminger06} previously employed by \cite{penza2006} to reproduce TSI variations. From 2005 to 2015, we used magnetic features area coverages obtained with the Precision Solar 
Photometric Telescope \citep[PSPT-][]{rast1999} located at the Mauna Loa Solar Observatory (MLSO), using the SRPM system \citep{fontenla2011,fontenla2005}. Both dataset were derived from full-disk CaIIK and red continuum images, 
but while the former distinguishes only between faculae and sunspots, the SRPM system provides area and position over the disk for features corresponding to the FAL atmosphere models utilized for the spectral synthesis.  
In order to homogenize the datasets, we associated the sum of the areas of models 1004 and 1005 to faculae $(A_{fac})$ and the sum of the area of models 1002 and 1003 to  network $(A_{net})$, moreover we took 
into account that from 1988 to 2004 we have a spot coverage considered as inclusive of umbra and penumbra (i.e. as sum of the areas of models 1006 and 1007), while better reconstructions are obtained by keeping 
separate their contributions.  We were able to separate the contribution of umbra $(A_{um})$ and penumbra $(A_{penum})$ in the overall spot coverage $(A_{spot}=A_{um}+A_{penum})$, by using the relation $A_{penum}=1.6 \cdot A_{um}$, 
as derived from PSPT data. The network area prior to 2004,
which is not provided in the San Fernando database, was estimated by computing the correlation coefficient between the network area derived by the PSPT images and the Mg II index. 
%%%%%%%%%%%%%%%%%%%%%%%%%%%%%
%\begin{figure}
%    \includegraphics[width=8cm,height=7cm]{spot_facular_area.eps}
% \includegraphics[width=8cm,height=7cm]{area_fac.eps}
%  \includegraphics[width=8cm,height=7cm]{area_spot.eps}
%  \caption{Temporal variation of facular (left) and spot (right) area in part per millions (ppm). Black: data obtained from the San Fernando Observatory. Red: data obtained from PSPT-MLSO.DELETE THESE FIGS}
%  \label{spot_facula_area}
%\end{figure}
%%%%%%%%%%%%%%%%%%%%%%%%%%%%%

We synthetized in this way the daily fluxes in the spectral ranges of interest, for each of the days in which the coverage of magnetic features were available. 
The values were then converted into irradiance and integrated in the FUV and MUV ranges. In the case of the Mg II h\&k lines, the spectra were convolved with a 1 nm Gaussian function and the Mg II index
was then computed following the formula in \citet{yeo2014}.

\section{Results} \label{sec:result}
%\textbf{We need to verify in the following how the data were binned and/or the temporal cadence. Daily data? monthly data? Vale, Mija, can you provide these info, please? Do you think we should smooth results in the plots?}

%We first investigated the impact on our reconstructions of using magnetic features area coverages derived from different instrument. To this aim, the spectral synthesis was extended to \textbf{how many microns?} and then integrated to estimates the Total Solar Irradiance. Variations of estimated TSI variations are compared to NOAA TSI composite (before 2004)  and SORCE-TIM from 2004 on (both datasets are available at \footnote{WEBSITE}) in Fig. \ref{TSI_rec} \textbf{QUESTE reference ai dataset sono giusti?} thus suggesting that our approach is sound. 

In this section we compare the reconstructed spectral irradiance temporal variations with variations obtained  from measurements. In particular, we compare the relative variation 
$\Delta \mathscr{F}(t)= [\mathscr{F}(t) -\mathscr{F}(t_{min.})]/<\mathscr{F}(t_{min.})>$, where $\mathscr{F}(t)$ is the spectral irradiance or index at time $t$ and $<\mathscr{F}(t_{min.})>$ is the monthly average during 
a period of minimum of activity, namely April 2009.

\begin{figure}
\centering
    \includegraphics[width=8cm,height=6.8cm]{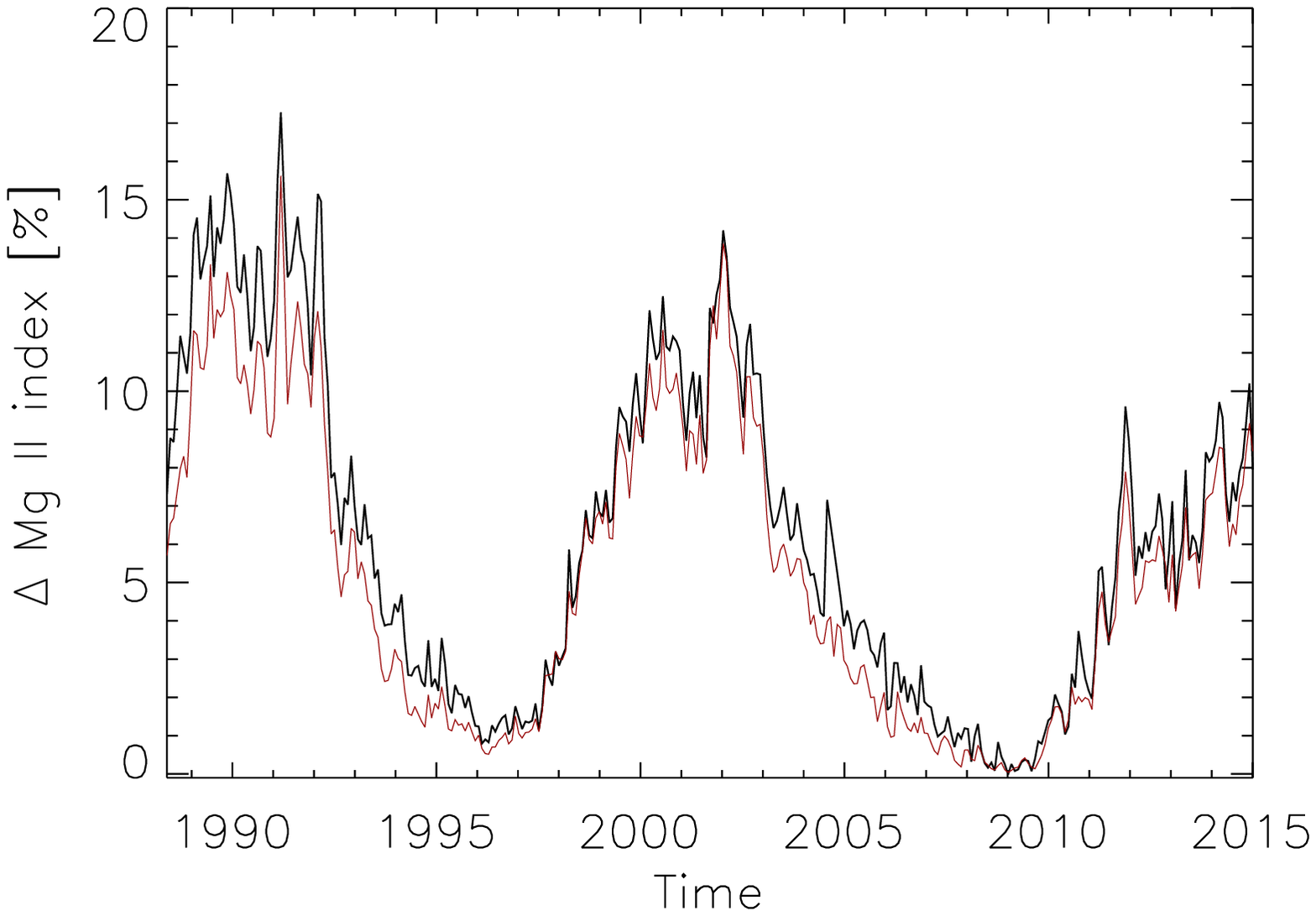} \\
    \includegraphics[width=8cm,height=6.8cm]{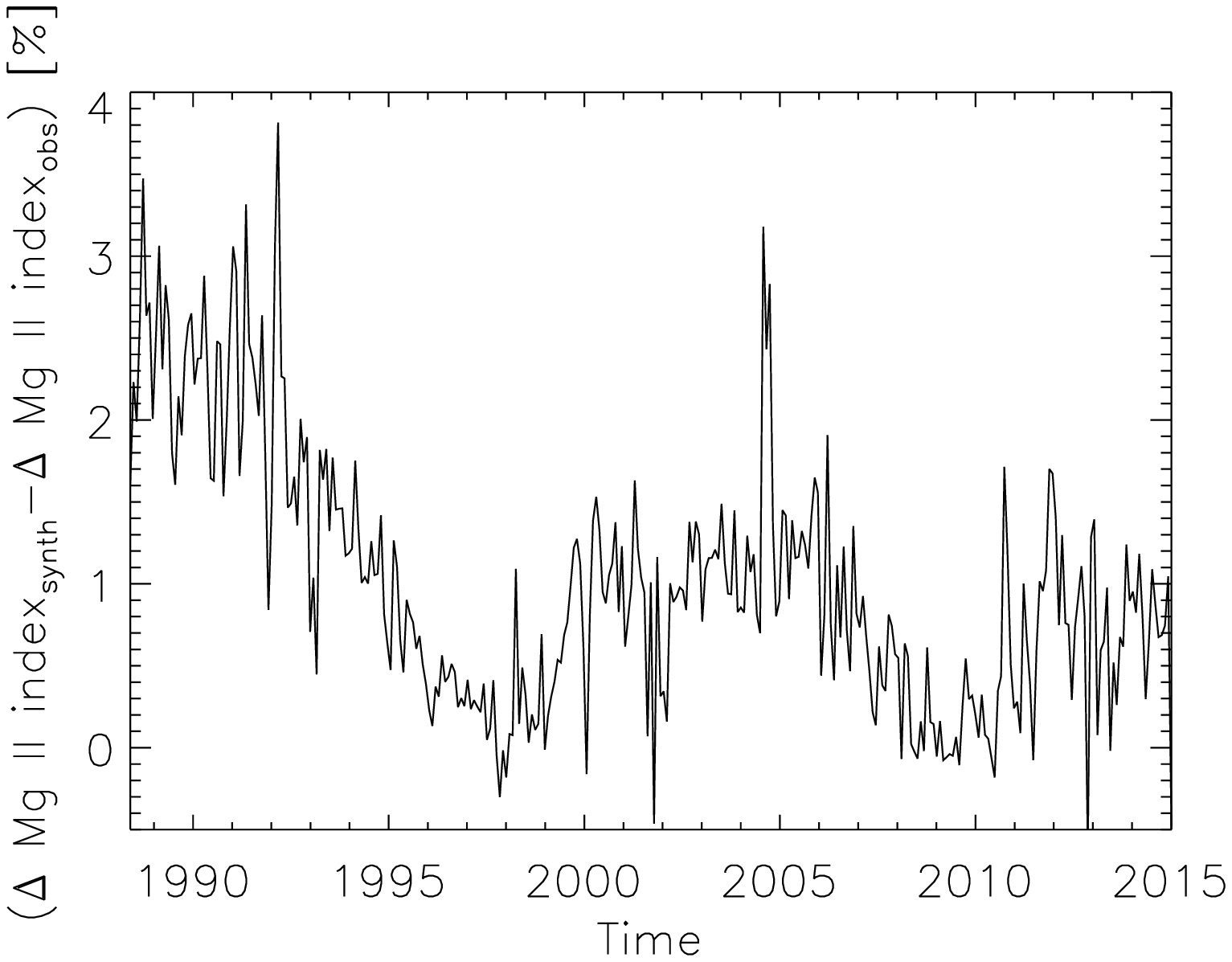}
  \caption{Top: reconstruction (black) of the Mg II index monthly variations from 1988 to 2015 compared to the Bremen composite variations (red). Bottom: difference between the synthetic and measured Mg II index variability. }
  \label{Mg II_rec}
\end{figure}

Reconstructions between April 1988 and January 2015 of monthly relative variations of the Mg II index are shown in Fig.~\ref{Mg II_rec} together with the Bremen index composite variability.   %\textbf{we decided to show daily variations, but for consistency with what showed later, we would use monthly averages?)}$. 
The agreement between synthetic and measured Mg II index variability is very good over the whole period analyzed, for which the Pearson's correlation coefficient is 0.98. 
The largest difference between measurements and reconstruction is found in 1992, during the maximum of Cycle 22, for which the Mg II index variability is overestimated up to approximately 4.\%. 

\begin{figure*}
    \includegraphics[width=8cm,height=6.8cm]{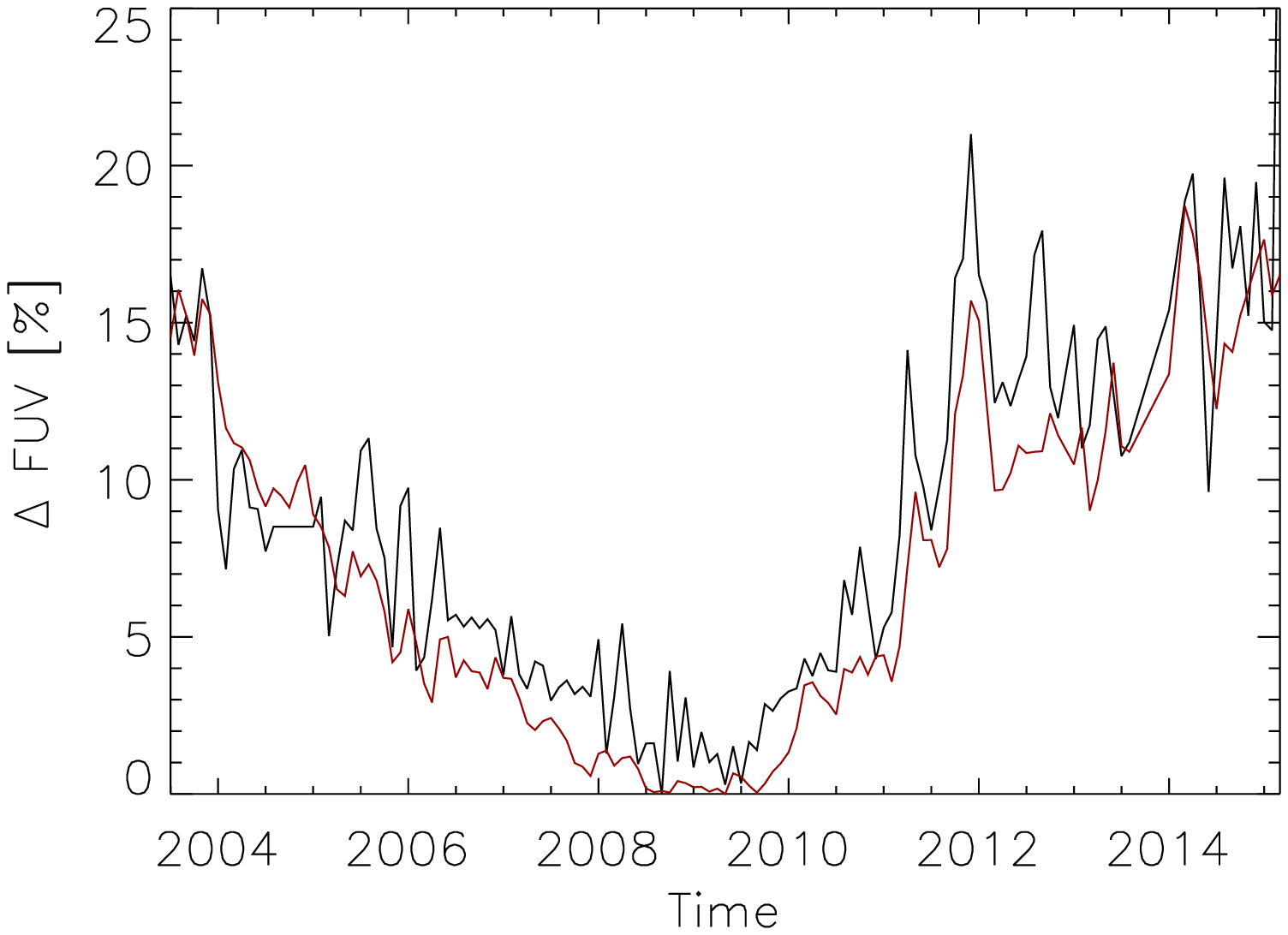}
    \includegraphics[width=8cm,height=6.8cm]{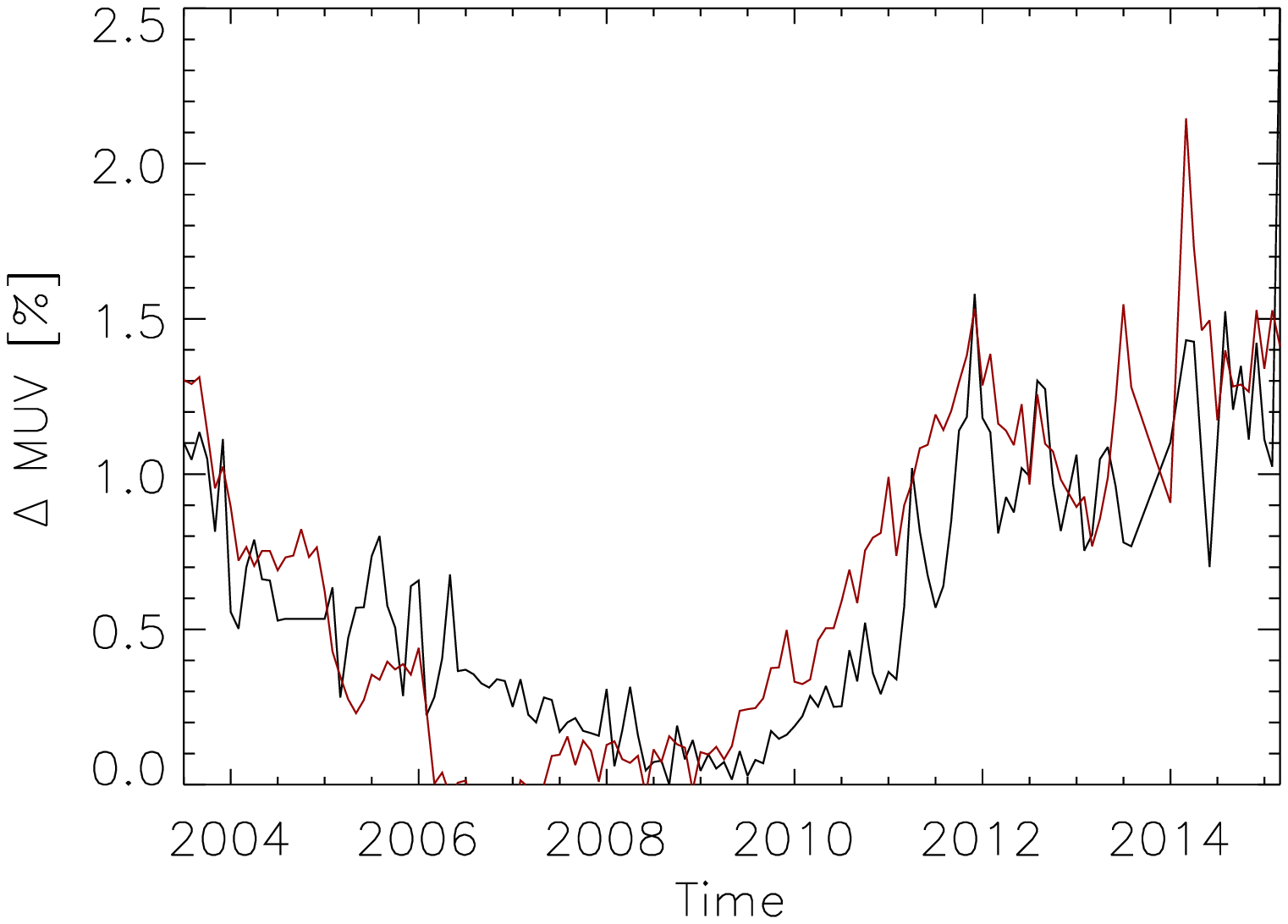}\\
    \includegraphics[width=8cm,height=6.8cm]{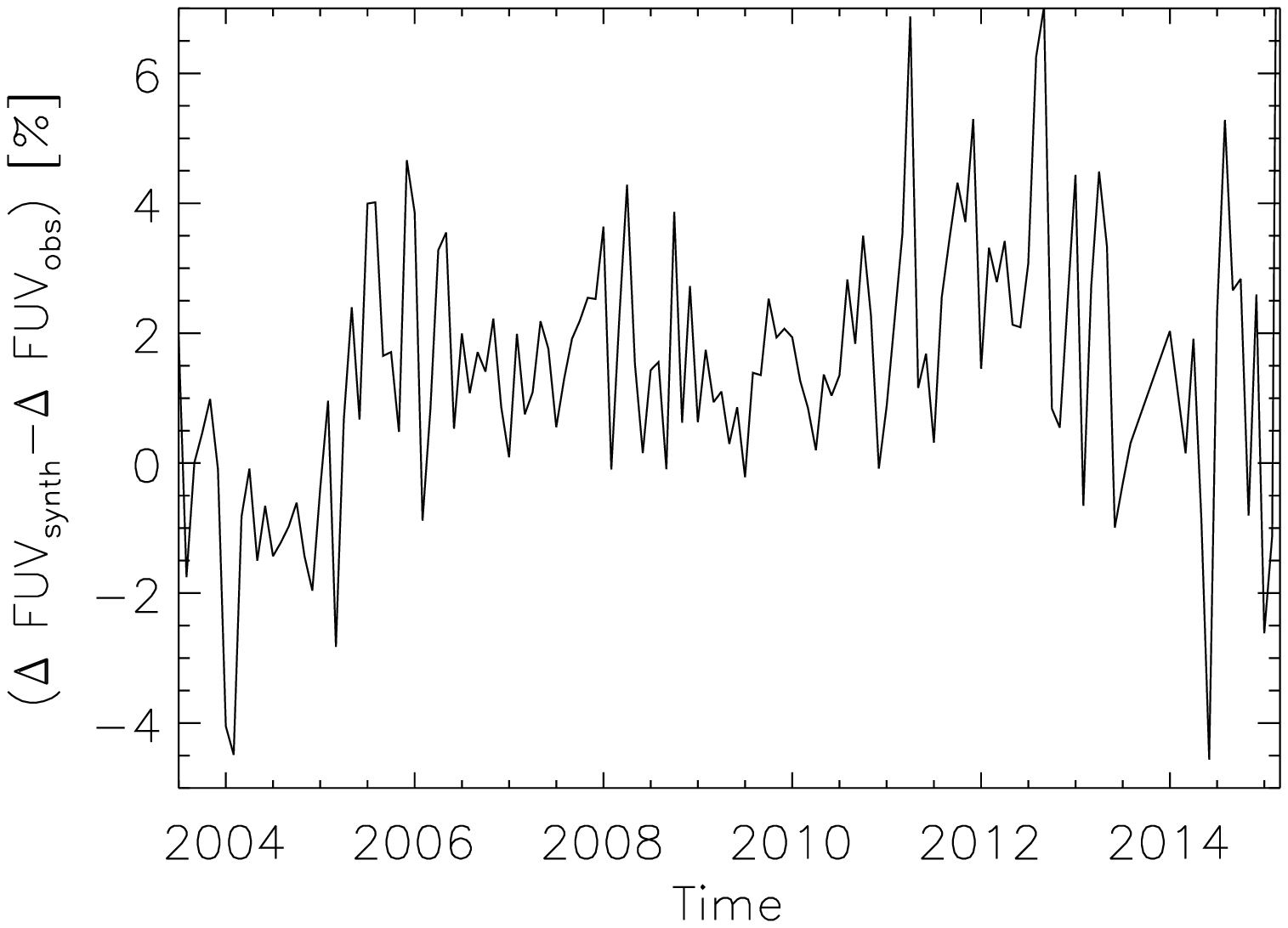}
    \includegraphics[width=8cm,height=6.8cm]{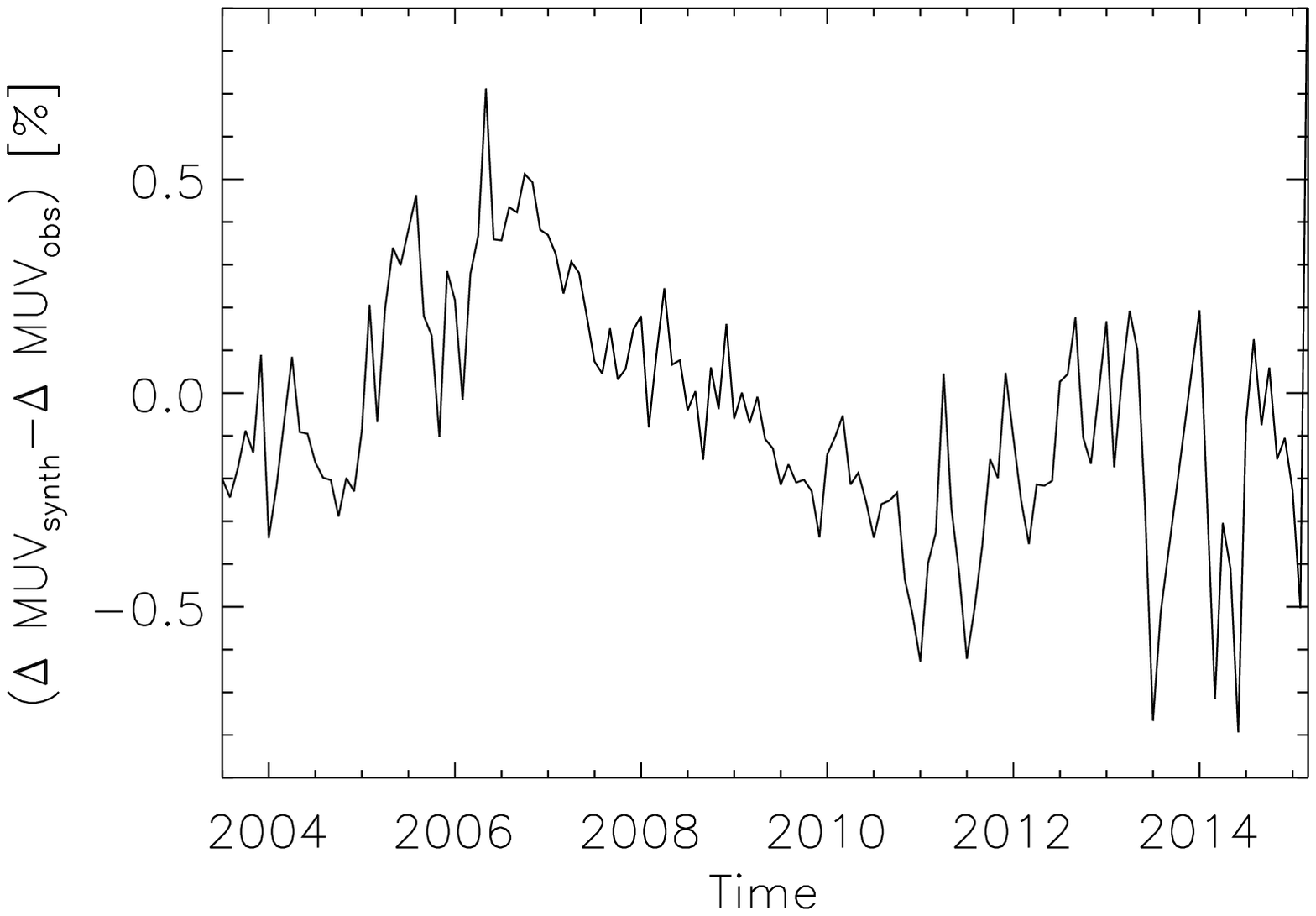}
  \caption{Top: monthly averages of variability obtained from synthesis (black) and SOLSTICE observations (red) in the integrated FUV (left) and MUV (right) spectral ranges. Bottom: differences between synthetic and observed FUV
  (left) and MUV (right) variability. }
  \label{MUV_FUV_rec}
\end{figure*}

Reconstructions of monthly variations in the FUV and MUV spectral ranges are shown in Fig.~\ref{MUV_FUV_rec}. In this case, comparison of synthetic data with observations are limited to the time in which SOLSTICE 
data became available, that is after April 2004. We note that overall the agreement between synthetic and observed variability is
better in the FUV range than in the MUV.  Indeed, the Pearson's correlation coefficients 
between the observed and reconstructed variability is 0.815 and 0.94 for the MUV and FUV spectral ranges, respectively. In the case of the FUV, the variability is consistently overestimated, 
with maximum difference  between observed and synthetic data being of about 6\% between 2012 and 2013. In the case of the MUV, our synthesis consistently underestimates the variability, with a 
maximum difference of about 0.8\%  found in 2014 data. 

%%%%%%%%%%%%%%%%%%%%%%%%%%%%%
%%%%%%%%%%%%%%%%%%%%%%%%%%%%%

%%%%%%%%%%%%%%%%%%%%%%%%%%%%%
\begin{figure*}
\centering{
\includegraphics[width=8.5cm,height=7.5cm]{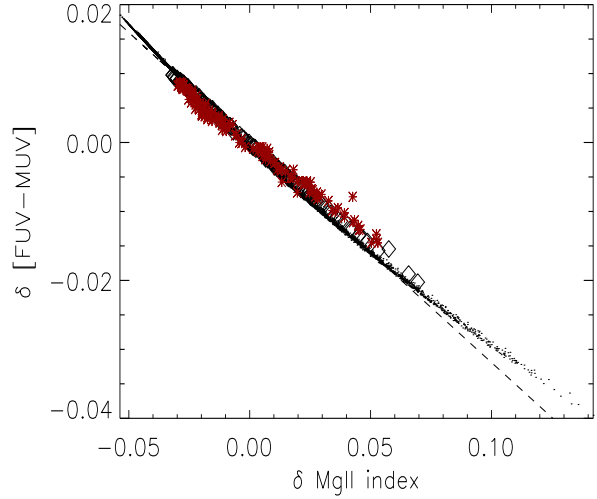}
  \caption{Correlations between observed and reconstructed [FUV-MUV] color and Mg II index. Black dots: indices reconstructed over the temporal range 1988-2015. Red stars:  indices computed from SORCE data over the temporal 
  range 2004-2015. Black diamonds: indices reconstructed and interpolated over the same temporal range as SOLSTICE observations. Note that variations have been computed with respect 
  to the average value over the indicated temporal range (see text). The dashed line represents the linear fit obtained from reconstructed data over the period 1988-2015.}
  \label{COLOR_vs_Mg II}
  }
\end{figure*}

%%%%%%%%%%%%%%%%%%%%%%%%%%%%%
We finally show the dependence of relative variations of the [FUV-MUV] Color Index on relative variations of the Mg II index in Fig.~\ref{COLOR_vs_Mg II}, as obtained by our reconstructions and SOLSTICE measurements.
Following \citet{lovric2017} the scatter plot shows monthly averaged data and the 
differences are computed with respect to the average value over the period analyzed, that is\\ $\delta \mathscr{F}(t)= [\mathscr{F}(t) -<\mathscr{F}(t)>)]/<\mathscr{F}(t)>$ .

The different symbols in the plot show 
variations derived from our reconstruction and observations. To investigate to what extent the comparison is affected by the fact that reconstructions and measurements were derived from time series of different length and sampling,
we also show reconstructions computed interpolating the data at the same temporal interval and sampling of the SOLSTICE observations. The agreement between reconstructions and observations is very good, 
especially for reconstructions 
interpolated at the SOLSTICE temporal sampling.

The plot confirms that indices obtained from both, the observations  and synthetic reconstruction, are highly linearly correlated, the Spearman's correlation coefficient being in all three cases larger than 0.99.
These results support findings presented in \cite{lovric2017} that the [FUV-MUV] color index strongly correlates with the Mg II index, with the [FUV-MUV] color index decreasing as the solar activity increases.
%and the scenario of a variable chromosphere/photosphere where most relevant variations of the emission can be explained by a variable background, and are not particularly sensitive to the presence of emission lines. 
Moreover, the same assumptions made in the synthetic model of atmosphere lead to the exclusion of a dependence of the [FUV-MUV] color with the ascending and descending phase of the solar cycle and support the conclusion 
that UV instrumental degradation must be included in the analysis of the SOLSTICE FUV flux.

The comparison of results presented in Fig.~\ref{MUV_FUV_rec}  with those presented in Fig.~\ref{COLOR_vs_Mg II} suggests that the small differences between the observed and reconstructed FUV and MUV 
fluxes are somewhat compensated for in the computation of the [FUV-MUV] color index. On the other hand, the small deviation from linearity noticeable in Fig.~\ref{COLOR_vs_Mg II} at high values of Mg II index variations has
to be more likely imputed to overestimation of the reconstructed Mg II index variability, as shown in Fig.~\ref{Mg II_rec}.

%%%%%%%%%%%%%%%%%%%%%%%%%%%%%
%%%%%%%%%%%%%%%%%%%%%%%%%%%%%

\section{Discussion} \label{sec:disc}
This paper presents  synthetic reconstructions of the [FUV-MUV] color and Mg II index  that well reproduce observations obtained with SORCE/SOLSTICE over the period May 2003 - January 2015.  
The presented reconstruction  method combines spectra obtained with semi-empirical  atmosphere models and area coverage of magnetic features derived from the analysis of full-disk images. 
Our approach is similar to other semi-empirical approaches previously presented in the literature, such as SATIRE \citep{unruh2000,krivova2006}, COSI \citep{shapiro2010}, 
SRPM \citep{fontenla2011,fontenla2015} and OAR \citep{penza2003, penza2006, ermolli2011}. 

%The UV irradiance variability estimated with our reconstruction agrees very well with SORCE/SOLSTICE observations. Although we plan to investigate a detailed comparison with variability obtained with different instruments and 
%different reconstruction techniques in a future work, we note that our reconstruction seems to produce a better agreement with SORCE/SOLSTICE measurements than what has been previously obtained with other
%reconstruction techniques \citep[e.g.][]{yeo2014b,yeo2017}.
%%%%%%%%%%%%%%%%%%%%%%%%%%%%%
\begin{figure*}
\centering{
    \includegraphics[width=7cm,height=6.3cm]{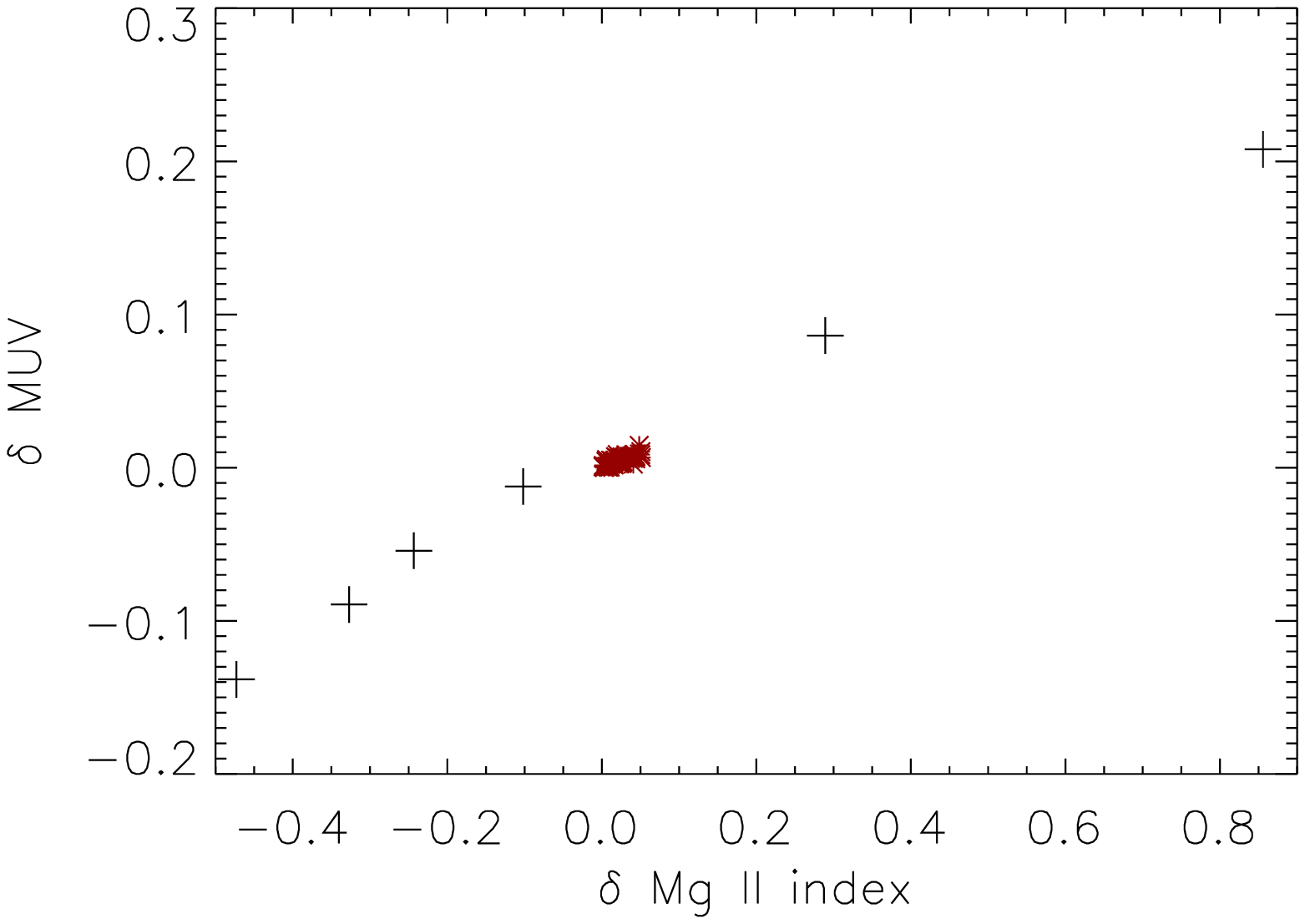}
    \includegraphics[width=7cm,height=6.3cm]{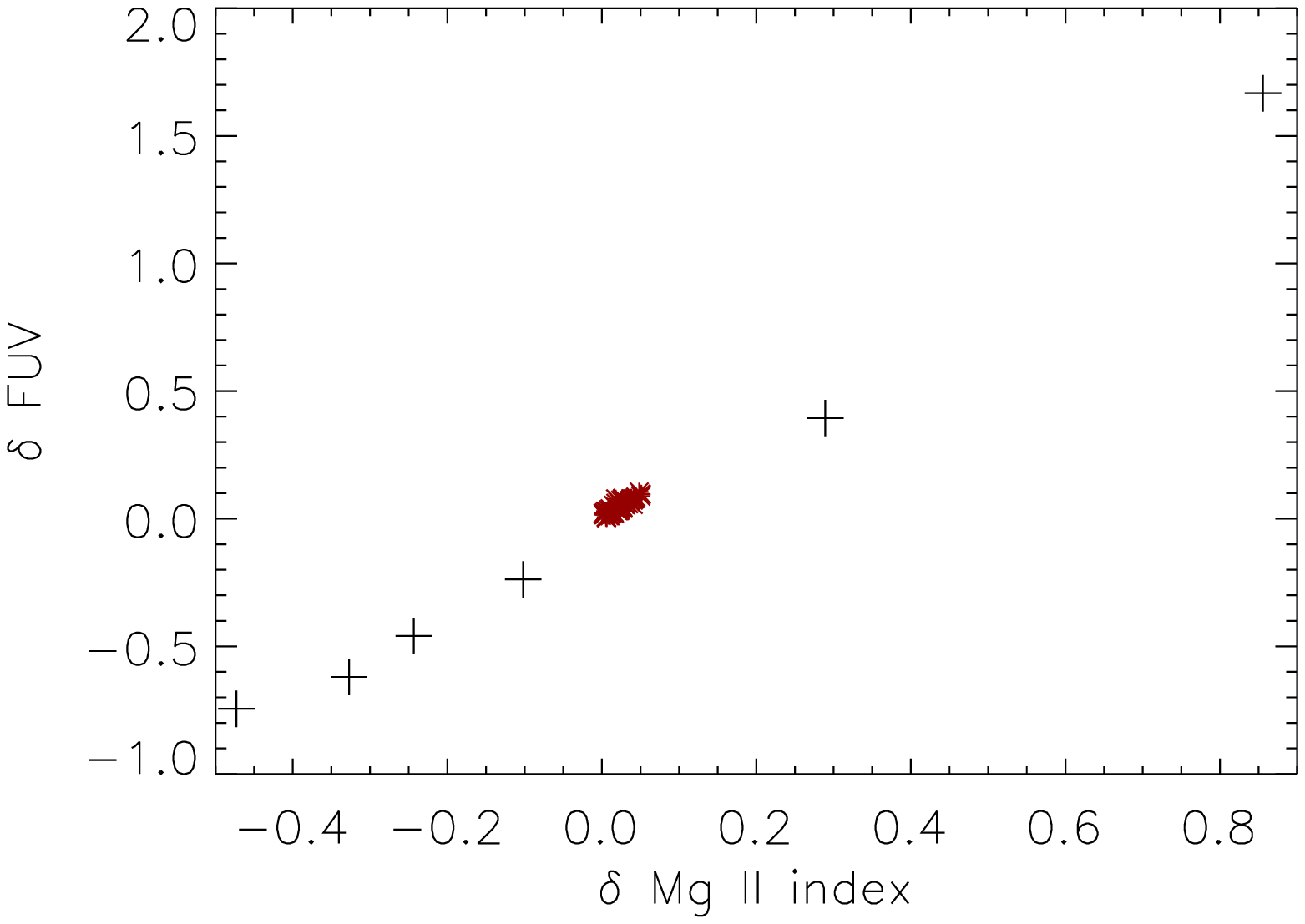}
 \caption{ Variations of the  MUV (left) and FUV (right) versus variations of the Mg II index as derived by FAL2011 models (plus) and SOLSTICE observations (red stars). Only FAL models representing quiet, 
 network and facular features (from model 1000 to model 1005) are shown in the plots.}
  \label{MUVFUV_FAL}
  }
\end{figure*}

A very important indication, in our opinion, that our synthesis reproduces correctly the main properties of the FUV and MUV regions of the solar spectrum is supported  by the results reported in Fig.~\ref{MUVFUV_FAL}, 
which show the  correlation between lambda-integrated MUV and FUV values and Mg II index computed by using different FAL2011 models (i.e. by supposing  a sun completely quiet, a sun completely covered 
with network, etc.) together with the  values measured with SOLSTICE along the solar cycle. The plot highlights that the "real" Sun  is located in the "middle" of the network  (model 1003) and facula 
(model 1004) models. This result is in 
agreement with the finding by \citet{linsky2012} and \citet{linsky2014} that the measured EUV and FUV continuous emission from  stars of various level of activity well correlate with synthetic spectra obtained 
with FAL sets of models \citep{fontenla2011,fontenla2015} using an NLTE synthesis. 
In other words, although it is established that the UV radiation is not reproducible in LTE, results presented in this paper suggest that the UV irradiance variability, at least in the FUV and MUV ranges, can be well 
reproduced considering a simple LTE approach. This is not entirely surprising for the MUV, for which, as already discussed in detail in Sec.~\ref{sec:atmos}, LTE synthesis have been already employed in other 
reconstruction techniques, but is rather unexpected for the FUV, which is flecked by a number of emission lines not reproduced by our synthesis, especially at wavelengths shorter than approximately 150 nm 
(see Fig.~\ref{MUVFUV_vs_WHI}). In this respect it is important to note that wavelengths longer than 150 nm contribute for more than 65\% to the whole FUV even during periods of high activity.   
%We believe the satisfying agreement with observations has to be attributed to the fact that the integrated value of the irradiance in this spectral range is dominated by radiation at wavelengths longer 
%than approximately 150 nm, where our synthesis shows a good agreement with the WHI reference spectrum. 

The strong linearity between the two indices shown in Fig.~\ref{COLOR_vs_Mg II} must be ascribed to their strong temperature dependence and to the proximity of the Mg II core and FUV formation heights on one side, 
and Mg II continuum and MUV on the other. 
Most of the radiation contributing to the FUV and MUV  originates in the higher photosphere and chromosphere \citep{thuillier2012}, where
atmosphere models of quiet and magnetic structures present similar (and rather small) temperature and density 
gradients \citep[c.f.r. Fig.~1 in ][]{fontenla2011}. This means that, at least at first order, an increase of the activity corresponds to an increase of equal amount of the temperature and density from which both MUV and FUV,
as well as Mg II wings and core, originate.
Under the assumption that the lambda-integrated fluxes can be described by a Plank function, it easy to show that the color index must scale linearly with the temperature, at least for small temperature variations.
Therefore, since the spectral brightness variations
over a magnetic cycle do not exceed few tens of K \citep{fontenla2011}, the [FUV-MUV] color index must also scale linearly with the activity. In LTE, a linear dependence of the core/wing ratio is expected
\citep[e.g.][]{criscuoli2013}, but such dependence is probably less obvious for the Mg II h\&k line, whose cores form in NLTE conditions. Nevertheless, collisions play an important role in the formation of the line \citep{linsky1970},
making the core of this doublet strongly sensitive 
to temperature \citep[e.g.][]{carlsson2015}. Indeed, numerical magneto hydrodynamic simulations indicate that the brightness temperature derived from the Mg II h\&k cores is a good approximation for the plasma temperature at 
the corresponding formation heights \citep{leenaarts2013b}. Moreover,
the Bremen Mg II index is derived from data of relatively modest spectral resolution, so that the cores are not well resolved. This means that the core intensities used to construct the Mg II index must have a 
non-negligible contribution from several layers of the atmosphere, 
from the lower photosphere, where the response function of the spectral region between the two h\&k peaks reaches its maximum \citep{uitenbroek1997}, to the base of the transition region \citep{thuillier2012,leenaarts2013b,carlsson2015},
where the cores form. 
Employing again a Plank function to describe both, the core and the wing intensities of the Mg II doublet, it is easy to show that the Mg II index must scale linearly with temperature, at least for small temperature perturbations.
%The dependence on the chromospheric temperature of the 
%various quantities analyzed is illustrated in Fig.\ref{vstemp}. Here the chromospheric temperature is computed as the temperature corresponding to optical depth $\tau_{500} = 0.8\cdot 10^{-6}$, which, for the quiet sun model 1001 corresponds to 
%about 1170 km above the photosphere. A linear dependence of the FUV and MUV irradiance and of the Mg II index on the chromospheric temperature is found for models from 1000 to 1003, corresponding to quiet and network features,
%while for the  two facular models
%(1004 and 1005)  the relation deviates from being linear. Because quiet and network regions are the most abundant features on the solar surface \citep{fontenla2011}, the [FUV-MUV] color and 
%Mg II index also scale linearly with the activity.     

\section{Conclusions} \label{sec:concl}
Results produced by the solar UV irradiance reconstructions presented in this paper reproduce with an excellent agreement the SORCE/SOLSTICE data trends compensated for instrumental 
degradation according to the method describe in \citet{lovric2017}. This suggests that the proposed technique does not overcompensate irradiance variations and, at the moment, rules out possible
temporal variations of the [FUV-MUV] color - Mg II index relation, at least during the time range analyzed, confirming  that the [FUV-MUV] color index strongly correlates with the Mg II index, 
with the UV color index decreasing as the magnetic activity increases. 

\section{Acknowledgments}
The National Solar Observatory is operated by the Association of Universities for Research in Astronomy, Inc. (AURA) under cooperative agreement with the National Science Foundation.
This work was partially supported by the Joint Research PhD Program in “Astronomy, Astrophysics and Space Science” between the universities of Roma Tor Vergata, Roma Sapienza and INAF. The authors are grateful to Dr. Jerry Harder for providing
the PSPT masks, Dr. Dario del Moro for helping with the data format conversions and Dr. Han Uitenbroek for reading the paper and providing useful comments.
The following institutes are acknowledged for providing the data: Laboratory for Atmospheric and Space Physics (Boulder, CO) for SORCE SOLSTICE SSI data (http://lasp.colorado.edu/home/sorce/data/) and University of Bremen (Bremen, Germany) 
for Mg II index data (http://www.iup.uni-bremen.de/gome/gomemgii.html).  The authors are also thankful to the International Space Science Institute, Bern, for the support provided to the science team 335.  
\bibliography{sample} % your references Yourfile.bib
\end{document}